\documentclass[11pt,a4paper,epsfig]{article}

\usepackage{amsfonts}
\usepackage{graphicx}
\usepackage{cite}
\usepackage[colorlinks=true,linkcolor=blue]{hyperref}

\title{\bf Kinks in massive non-linear ${\mathbb S}^1\times{\mathbb S}^1$-Sigma models  }

\author{A. Alonso-Izquierdo$^{(a,b)}$, A.J. Balseyro Sebastian$^{(b)}$, J. Mateos Guilarte$^{(b)}$ and\\  M. A. Gonzalez Leon$^{(a,b)}$
\\ {\normalsize {\it $^{(a)}$ Departamento de Matematica
Aplicada}, {\it Universidad de Salamanca, SPAIN}}\\ {\normalsize {\it $^{(b)}$ IUFFyM}, {\it Universidad de Salamanca, SPAIN}}}

\date{}

\begin{document}

\maketitle

\begin{abstract}
In this paper the whole kink varieties arising in several massive non-linear Sigma models whose target space is the torus ${\mathbb S}^1\times{\mathbb S}^1$ are analytically calculated. This possibility underlies the construction of first-order differential equations by adapting the Bogomolny procedure to non-Euclidean spaces. Among the families of solutions non-topological kinks connecting the same vacuum are found. This class of solutions are usually considered to be not globally stable. However, in this context the topological constraints obtained by the non-simply connectedness of the target space turn these non-topological kinks into globally stable solutions. The analytical resolution of the equations allows the complete study of the linear stability for some basic kinks.
\end{abstract}

%PACS:

\section{Introduction}

Topological defects have been successfully recognized over the last fifty years as the source of a wide variety of non-linear phenomena arising in different physical frameworks, including Optics \cite{Mollenauer2006,Schneider2004,Agrawall1995}, Condensed Matter \cite{Bishop1980,Eschenfelder1981,Jona1993,Strukov}, Molecular systems \cite{Davydov1985,Bazeia1999}, Biochemistry \cite{Yakushevich2004}, Cosmology \cite{Kolb1990,Kibble1976,Vilenkin1994,Vachaspati2006}, etc. The most remarkable feature of these objects is that they behave as extended particles moving in the physical substrate because their energy densities are localized around one point. In general, topological defects are solutions in field theory models which cannot decay to the homogeneous ground state, the vacuum, because of topological constraints in the configuration space. The general pattern giving rise to topological defects arises when the configuration field space is the union of disconnected pieces associated to some spontaneous symmetry breaking. The emergence of this concept is due to the existence of degenerate manifolds of vacua. Topological defects correspond to spatially non-homogeneous ground states that only at the boundary of space tend to the homogeneous ones. If some homotopy group of the vacuum manifold is non-trivial the configuration space is formed by the union of several topologically disconnected pieces. In the particular case of scalar field theories defined on one spatial dimension space, the topological defects are solutions of non-linear partial differential equations connecting different homogeneous ground states at both ends of the spatial line. In this framework they are referred to as \textit{kinks} or \textit{solitons}. The $\phi^4$ and the sine-Gordon models are the paradigmatic relativistic examples in this context whereas the non-linear Schr$\ddot{\rm o}$dinger or Korteweg-de Vries equations are non-relativistic PDE's supporting solitons. The just mentioned above relativistic systems involve only one scalar field on a (1+1)-dimensional Minkowski space-time. The ubiquity of these two models in the physical literature is not only explained by the universality of some properties of the topological defects but also because analytical expressions characterizing the kink solutions can be found in these theories. The linear stability can also be studied in these models identifying the analytical form of the normal modes of vibration as well as their physical features. It is clear that the knowledge of exact expressions allows to carry out further analysis in the models leading to a deeper understanding of the previously mentioned non-linear phenomena. As a consequence the analytical identification of new kink solutions in diverse scalar field theories have become an active research topic for decades. For example, the so called \textit{deformation method} developed by Bazeia et al. is a technique that allows to construct exact kink solutions for new models derived from other already known cases, see references \cite{Bazeia2006a, Bazeia2006b, Almeida2004,Bazeia2006c,Souza2009, Cruz2009, Chumbes2010, Bazeia2011}. An important qualitative step in this issue corresponds to the move from one-component scalar field models to two-component scalar field models, passing from the integration of a single partial differential equation to the integration of a system of two partial differential equations \cite{Rajaraman1982}. In this context, it is worthwhile mentioning that two-parametric families of kinks have been identified for distinct models leading to a very rich structure of topological defects, see \cite{Bazeia1995, Shifman1998, Ito1985, Alonso1998, Alonso2002, Afonso2007, Afonso2008, Afonso2009, Alonso2013}. In these references, either Hamilton-Jacobi separability of the field equations characterizing static solutions or the Bogomolny arrangement of the field theory energy functional or new adapted deformation methods are usually employed to obtain the exact solutions. These techniques have also been used to obtain kink families in scalar field theories of three or more components \cite{Alonso2000, Alonso2004b}.

In the previously mentioned models, the $N$-component scalar field is a map from the Minkowski spacetime $\mathbb{R}^{1,1}$ to $\mathbb{R}^N$. However, some physical applications require the use of non-linear Sigma field theory models with different target spaces. This is, for example, the scenario found in spintronics, where the fields in the effective theory describe the continuous limit of spin chains in magnetic materials. In this framework the target manifold is the sphere ${\mathbb S}^2$, that is, a three-component scalar field must comply with the constraint $\phi^1 \phi^1+\phi^2\phi^2+\phi^3 \phi^3=R^2$, such that $\phi\equiv (\phi^1,\phi^2,\phi^3) : \mathbb{R}^{1,1} \rightarrow S^2$. For example, Haldane constructed a $O(3)$ non-linear sigma field theory model to describe the low-energy dynamics of large-spin one-dimensional Heisenberg antiferromagnets, see \cite{Haldane1983}. Distinct features of these materials, such as exchange, anisotropy or dipole-dipole interactions can lead to the existence of several non-collinear ground state configurations in these substances. This behavior can give place to the emergence of topological defects. Indeed, Haldane semiclassically quantizes the kink type solution of the model. Despite the fact that the analytical identification of topological defects in this class of non-linear $S^2$-Sigma models is a more challenging problem than its linear partners some advances have been achieved in this direction. For example, in \cite{Alonso2008b, Alonso2010} exact expressions of the spin solitary waves in a similar model than that addressed by Haldane are identified.  Other relevant work in this spirit is the identification of chiral magnetic soliton lattices present on a chiral helimagnet ${\rm Cr}_{\frac{1}{3}}{\rm NbS}_2$ \cite{Togawa2012}. Magnetic topological defects with applications to logical operations and/or information storage have been also investigated in \cite{Koumpouras2016}. On the other hand, the existence of these topological solutions in supersymmetric massive non-linear sigma models have also been profusely studied, see \cite{Eto2006, Arai2002, Dorey1998,Naganuma2001b}.

Another new interesting qualitative step in this research line is the introduction in the non-linear Sigma models of non-simply connected target spaces. This is the scenario which will be addressed in this paper. Thus, the main concern in this paper is to replace the target space of the non-linear sigma models described in \cite{Haldane1983}, \cite{Alonso2008b}, \cite{Alonso2008c} and \cite{Alonso2010} which is a Riemann surface of genus zero by another Riemann surface but in the present case of genus one. In precise detail, we shall search for analytical expressions of kinks solutions in ${\mathbb S}^1\times{\mathbb S}^1$-Sigma models, where the target space corresponds to a two-dimensional torus. This study has an intrinsic mathematical interest with physical repercussions. For example, non-topological kinks (which are characterized by the fact that they asymptotically begin and end at the same vacuum points) are usually considered unstable because the elastic forces derived from the spatial derivatives in the action functional make the kink orbit be continuously shrunk to the vacuum point. In other words, the non-topological kink decays to the vacuum. However, this premise can be eluded if the target space is non-simply connected. Indeed, this is the case found in this paper for some non-topological kinks on the torus, where the linear stability analysis is also implemented to corroborate the previous claim. On the other hand, toroidal surfaces play an essential role in different physical applications. For instance, in recent years the presence of disclinations in toroidal nematics have been extensively investigated, see \cite{Evans1995, Selinger2011, Ye2016, Li2014}. For liquid crystals in the nematic phase a disclination is a line defect solution characterized by the fact that the rotational symmetry is spontaneously broken making the nearby rod-shape organic molecules self-align in a parallel way. The effective field theories describe the behavior of the continuum limit of the nematic director field. In this scenario the target space is the tangent bundle of the torus. Solutions in these models are numerically identified. Another different framework is given by membrane solitonic solutions arising in M-theory. In particular, in \cite{Axenides2002,Brugues2005, Alvarez2022} spinning membranes toroidally compactified on $M_9\times \mathbb{T}^2$ are discussed. In this case, the torus is trivially embedded in the target space. The present paper is intended to bring further analytical insight in this context. As previously mentioned the purpose of this work is the search of kink solutions in non-linear $(\mathbb{S}^1 \times \mathbb{S}^1)$-Sigma models and study their linear stability.

The organization of this paper is as follows: in Section \ref{sec:2} the general theoretical background of non-linear Sigma models on the torus is introduced. Simple toroidal angular coordinates are used to describe the dynamics in this Section. A Bogomolny decomposition of the energy functional based on these new variables is considered, which leads to first order differential equations satisfied by the static kink solutions in this framework. In Section \ref{sec:3} a particular family of non-linear $(\mathbb{S}^1 \times \mathbb{S}^1)$-Sigma models is proposed. In particular the number of vacuum points depends on the values of an integer parameter $n_1$ and an integer or half-integer parameter $n_2$. The range of these parameters is chosen to guarantee the periodicity of the potential term on the torus. Particular models involving eight, four and two vacua are thoroughly studied in this section obtaining analytical expressions for the kinks. In general, some of the topological sectors are filled with dense families of kinks while others include only some isolated singular kinks. Sum rules between energies of these kinks are identified and analytically explained. When possible the linear stability is also examined. It is worthwhile mentioning that in the last model globally stable non-topological kinks can be found. Maybe, in this context despite the fact that these solutions asymptotically connect the same vacuum, the adjective \textit{non-topological} can be misleading because the simply-connectedness of the torus is the topological property which prevents the non-topological kinks from decaying into a vacuum point in this class of theories. However, in this paper we shall continue using the \textit{non-topological} term to denote any solution with a closed orbit. Some conclusions are drawn in Section~\ref{sec:4}. Finally, some of the magnitudes used in this paper concerning the differential geometry on the torus are explicitly written in Appendix.

\section{Non-linear $(\mathbb{S}^1 \times \mathbb{S}^1)$-Sigma models in (1+1)-dimensions}

\label{sec:2}

We shall deal with $(1+1)$-dimensional nonlinear Sigma models whose target manifold is given by the torus $\mathbb{S}^1 \times \mathbb{S}^1$ embedded in the Euclidean space $\mathbb{R}^3$. The Minkowski metric $\eta_{\mu\nu}$ has been chosen as $\eta_{00}=-\eta_{11}=1$ and $\eta_{12}=\eta_{21}=0$. The real scalar fields $\phi^i: \mathbb{R}^{1,1} \rightarrow \mathbb{R}$, $i=1,2,3$, involved in this theory must comply with the constraint
\begin{equation}
f(\phi^1,\phi^2,\phi^3) \equiv \Big(R-\sqrt{\phi^1 \phi^1 + \phi^3\phi^3}\Big)^2 + \phi^2\phi^2 -r^2=0 \, ,
\label{constraint}
\end{equation}
where $r$ is the radius of the torus tube (minor radius) and $R$ is the distance from the center of the tube to the center of the torus (major radius). The condition
\[
R>r>0
\]
is imposed to avoid self-intersections in the target space. We thus deal with the set of maps from the Minkowski space-time to a genus $1$ Riemann surface target space:
\begin{equation}\vec{\Phi}: \mathbb{R}^{1,1} \quad \longrightarrow \quad \mathbb{T}^2\equiv \mathbb{S}_R^1\times \mathbb{S}_r^1 \nonumber \,
\end{equation}
because the embedded surface in ${\mathbb R}^3$ determined by the algebraic equation (\ref{constraint}) is the direct product of a circle of radius $R$ times another circle of radius $r$. The dynamics in this type of models is governed by the action functional
\begin{equation}
S\left[\phi\right]=\displaystyle \int_{\mathbb{R}^{1,1}} \left[\frac{1}{2}\displaystyle\sum_{i=1}^3 \left(\frac{\partial \phi^i}{\partial t}\right)^2- \frac{1}{2}\displaystyle\sum_{i=1}^3 \left(\frac{\partial \phi^i}{\partial x}\right)^2 - V(\phi) - \lambda f(\phi)\right] ~ dx ~ dt \, ,
\label{action}
\end{equation}
where $V:\mathbb{R}^3 \rightarrow \mathbb{R}$ is a non-negative \textit{potential function} determining every particular model and $\lambda$ is the Lagrange multiplier associated with the constraint (\ref{constraint}). Note that we use $\phi=(\phi^1,\phi^2,\phi^3)$ in (\ref{action}) in order to alleviate the notation. The pure non-linear ${\mathbb S}^1\times{\mathbb S}^1$-Sigma model corresponds to the choice: $V(\phi)=0$.  

The torus parametrization
\begin{eqnarray}
\phi^1(\theta,\varphi) &=& (R+r\sin \theta)\cos \varphi \, ,  \nonumber \\
\phi^2(\theta,\varphi) &=& r \cos \theta \label{tor} \, ,\\
\phi^3(\theta,\varphi) &=& (R+r\sin \theta)\sin \varphi \nonumber
\end{eqnarray}
on angular variables $\theta\in [0,2\pi)$ and $\varphi\in[0,2\pi)$, or equivalently:
\begin{equation}
\theta(\phi) = \arctan \frac{\sqrt{(\phi^1)^2+(\phi^3)^2}-R}{\phi^2} \, ,\hspace{1cm}
\varphi(\phi) = \arctan \frac{\phi^3}{\phi^1} \, , \label{inverse}
\end{equation}
allows us to comply with the constraint (\ref{constraint}) in an implicit way. The use of the angular variables (\ref{tor}) in (\ref{action}) leads to the action functional
\begin{equation}
S[\theta, \varphi]= \int_{\mathbb{R}^{1,1}}  \, \left[ \frac{1}{2} r^2 \eta^{\mu \nu}\frac{\partial\theta}{\partial x^\nu}\frac{\partial\theta}{\partial x^\mu}+\frac{1}{2}(R+r\sin\theta)^2 \eta^{\mu \nu}\frac{\partial\varphi}{\partial x^\nu}\frac{\partial\varphi}{\partial x^\mu} -V(\theta,\varphi) \right] \, dx ~ dt \, , \label{action2}
\end{equation}
where Einstein summation convention is employed for the space-time indices and the diagonal components of the metric tensor induced on the target space
\[
g= {\rm diag}\,\{r^2,(R+r\sin\theta)^2 \} \hspace{0.5cm}
\]
are involved, see Appendix. Notice that now $V(\theta,\varphi)$ denotes the restriction of the potential function to the torus. The Euler-Lagrange equations derived from (\ref{action2}) lead us to the non-linear PDEs
\begin{equation}
\eta^{\mu\nu} \frac{\partial^2 u^i}{\partial x^\mu  \partial x^\nu} + \eta^{\mu\nu}\Gamma_{jk}^i \frac{\partial u^j}{\partial x^\mu} \frac{\partial u^k}{\partial x^\nu}+ g^{ij} \frac{\partial V}{\partial u^j}=0  \, , \quad u^1=\theta \, ,\, u^2=\varphi \, \, \, , \, \, \, i,j=1,2 \, , \label{fenlv2}
\end{equation}
where Einstein summation convention is used for both latin and greek indices and the Christoffel symbols are detailed in the Appendix. The spacetime translation symmetry in the action functional (\ref{action2}) implies the conservation of the total energy
\begin{eqnarray*}
E[\theta, \varphi]&=& \int^{\infty}_{-\infty}  \, \left\{\frac{1}{2} r^2 \left[\left(\frac{\partial\theta}{\partial t}\right)^2+\left(\frac{\partial\theta}{\partial x}\right)^2\right]+\frac{1}{2}(R+r\sin\theta)^2 \left[\left(\frac{\partial\varphi}{\partial t}\right)^2+\left(\frac{\partial\varphi}{\partial x}\right)^2\right] +V\left[\theta,\varphi\right]\right\} \, dx
\end{eqnarray*}
for any solution $\Sigma(t,x) \equiv (\theta(t,x),\varphi(t,x))$ of the equations (\ref{fenlv2}).
The configuration space ${\cal C}$ in this framework is defined as
\[
{\cal C} =\{ \Sigma(t,x) \in \mathbb{S}^1 \times \mathbb{S}^1 \, |\, \, E[\Sigma(t,x)] < + \infty \} \, .
\]
As a consequence, the elements of ${\cal C}$ must comply with the asymptotic conditions
\[
\lim_{x\rightarrow \pm \infty} \frac{\partial \Sigma(x,t)}{\partial t} = \lim_{x\rightarrow \pm \infty} \frac{\partial \Sigma(x,t)}{\partial x} =0 \, ,\hspace{1cm}  \lim_{x\rightarrow \pm \infty}  \Sigma(x,t) \in {\cal M} \, ,
\]
where ${\cal M}$ is the set of zeroes of the potential term $V(\theta,\varphi)$, also refered to as the vacua of the model:
\[
{\cal M} = \{v_j =(\theta_j,\varphi_j) \in \mathbb{S}^1 \times \mathbb{S}^1 \, |\, \,  V(\theta_j,\varphi_j) =0, \,\, j=1,2,\dots \}\, .
\]
In this paper we are interested in searching for traveling solutions which can be derived from the static field equations
\begin{eqnarray}
&& \frac{d^2 \theta}{dx^2} - \frac{1}{r} \cos \theta (R+r\sin \theta) \frac{d \varphi}{dx} \frac{d\varphi}{dx} - \frac{1}{r^2} \frac{\partial V}{\partial \theta}=0 \, , \label{new1b} \\
&& \frac{d^2 \varphi}{dx^2} + \frac{2r \cos \theta}{R+r\sin \theta} \frac{d \theta}{dx} \frac{d\varphi}{dx} - \frac{1}{(R+r\sin\theta)^2} \frac{\partial V}{\partial \varphi}=0\, , \label{new2b}
\end{eqnarray}
by applying a Lorentz boost. In particular, the set ${\cal M}$ of vacuum points
verify these requirements. They are the zero energy static homogeneous solutions of (\ref{fenlv2}). Our interest, however, focuses on static kink solutions, time-independent finite energy solutions of (\ref{fenlv2}) whose energy densities are localized and can be interpreted as extended particles in a physical substrate. In this framework, the potential term $V(\theta, \varphi)$ must be chosen as non-negative. Furthermore, in order to use the Bogomoln'nyi procedure in this context we shall restrict our study to models whose potential term $V$ can be written in the form
\begin{equation}
V(u^1,u^2)= \frac{1}{2} g^{ij}\, \frac{\partial W}{\partial u^i} \frac{\partial W}{\partial u^j} \, ,\hspace{1cm} u^1=\theta \, ,\, u^2=\varphi \, \, \, , \, \, \, i,j=1,2 \, ,  \label{ps}
\end{equation}
where $W:\mathbb{S}^1 \times\mathbb{S}^1\rightarrow\mathbb{R}$ is a differentiable function called superpotential. By applying the Bogomoln'nyi arrangement the energy functional reads
\[
E[\theta,\varphi]=\frac{1}{2}\int dx \,\left(g_{ij} \left[\frac{du^i}{dx}+ (-1)^a \, g^{mi} \frac{\partial W}{\partial u^m}\right] \left[\frac{du^j}{dx}+(-1)^a \, g^{nj} \frac{\partial W}{\partial u^n}\right]\right) + T \, ,
\]
where $a=0, 1$ and $T$ is defined as
\begin{equation}
T=\left\vert \int dx \,   \frac{du^i}{dx}\frac{\partial W}{\partial u^i} \, \,\right\vert=  \left\vert \int dx \,  \left[\frac{d\theta}{dx}\frac{\partial W}{\partial\theta}+\frac{d\varphi}{dx}\frac{\partial W}{\partial\varphi}\right]\, \,\right\vert \, . \label{charge}
\end{equation}
BPS kinks satisfy the system of first order differential equations
\begin{equation}
\frac{du^i}{dx}=(-1)^a \, g^{ij} \, \frac{\partial W}{\partial u^j} \, ,\hspace{1cm} u^1=\theta \, ,\, u^2=\varphi \, \, \, , \, \, \, i,j=1,2, \label{bpsb}
\end{equation}
minimizing the energy functional $E[\theta,\varphi]$ in the configuration space ${\cal C}$. Note that the global factor $(-1)^a$ could be absorbed in the spatial variable $x$, which means that kinks and antikinks are obtained for different values of $a$. If the superpotential $W(\theta,\varphi)$ and the BPS kinks are smooth functions then the magnitude $T$ defined in (\ref{charge}) becomes a topological charge
\begin{equation}\label{eq:EnergyBPS}
   T=\Big\vert \lim_{x\rightarrow \infty} W[\theta(x),\varphi(x)]- \lim_{x\rightarrow -\infty}W[\theta(x),\varphi(x)]\Big\vert \, ,
\end{equation}
which depends only on the vacua which are asymptotically connected by the BPS kinks. Moreover, $T$ is the total energy of these topological defects. Naturally, it can be checked that these BPS kinks are also solutions of the field equations (\ref{new1b})-(\ref{new2b}).

Lastly, the stability of these kinks must be analyzed. The study of the linear stability of a solution $\Sigma(x)=(\theta(x),\varphi(x))$, i.e. the analysis of small fluctuations around a kink, leads to the spectral problem associated to the operator:
\[
\Delta_\Sigma \eta= -\left( \nabla_{\Sigma^\prime}\nabla_{\Sigma^\prime}\eta+
R(\Sigma^\prime,\eta)\Sigma^\prime +\nabla_\eta {\rm grad}V\right)
\]
where the covariant derivative of $\eta(x)$ and the action of the
curvature tensor on $\eta(x)$ are:
\[
\nabla_{\Sigma^\prime} \eta=\left(\eta^{\prime
i}(x)+\Gamma^i_{jk}\eta^j u^{\prime
k}\right)\frac{\partial}{\partial u^i}\  ;\qquad
R(\Sigma^\prime,\eta)\Sigma^\prime=u^{\prime
i}\eta^j(x)u^{\prime
k}R^l_{ijk}\frac{\partial}{\partial u^l}.
\]
$u^1=u^1(x)=\theta(x)$, $u^2=u^2(x)=\varphi(x)$, and $\eta(x)$ is a vector-field along the kink solution $\Sigma(x)$ in the torus. The Hessian of the potential function, evaluated at $\Sigma(x)$, reads:
\[
\nabla_\eta {\rm grad}V=\eta^i\left(\frac{\partial^2
V}{\partial u^i\partial u^j}-\Gamma^k_{ij}\frac{\partial
V}{\partial u^k}\right)g^{jl}\frac{\partial}{\partial u^l}
\]
Particularizing the previous expressions to the case of the torus, see Appendix, the spectral equation for the second-order small fluctuation operator is written in matrix form as:
\begin{equation}
{\cal H}[\Sigma(x)] \, \Psi_n(x) = \omega_n^2 \Psi_n(x) \, , \label{spectralproblem}
\end{equation}
where ${\cal H}[\Sigma(x)]$ is a $(2\times 2)$-matrix operator, whose components are expressed as
\begin{eqnarray*}
{\cal H}_{11} &=&  -\frac{d^2 }{dx^2}+ \left[\cos^2{\theta}-\frac{\sin{\theta}(R + r \sin{\theta})}{r}\right] \left(\frac{d\varphi}{dx}\right)^2  +\frac{1}{r^2}  \frac{\partial^2 V}{\partial \theta^2}\, , \\
{\cal H}_{12} &=&  \frac{2 \cos{\theta} (R + r \sin{\theta})}{r} \frac{d\varphi}{dx} ~  \frac{d}{dx} +\frac{1}{r^2} \frac{\partial^2 V}{\partial \theta \partial \varphi} \, ,\\
{\cal H}_{21} &=&   -\frac{2 r \cos{\theta}}{R+r \sin{\theta}} \frac{d \varphi}{dx} \frac{d }{dx} + \frac{2 r (r+R \sin{\theta})}{(R+ r\sin{\theta})^2} \frac{d \varphi}{dx} \frac{d \theta}{dx}+\frac{1}{(R + r \sin{\theta})^2}  \frac{\partial^2 V}{\partial \theta \partial \varphi} -\frac{2 r \cos{\theta}}{(R + r \sin{\theta})^3} \frac{\partial V}{\partial \varphi} \, ,\\
{\cal H}_{22} &=& -\frac{d^2}{dx^2} -\frac{2 r \cos{\theta}}{R+r \sin{\theta}} \frac{d \theta}{dx} \frac{d }{dx}+ \frac{1}{(R + r \sin{\theta})^2}  \frac{\partial^2 V}{\partial \varphi^2}
\end{eqnarray*}
and $\Psi_n(x)=(\psi_n^1(x),\psi_n^2(x))^t$ stands for the two-component eigenfunctions. The lack of negative eigenvalues in the spectral problem (\ref{spectralproblem}) implies that the solution $\Sigma(x)=(\theta(x),\varphi(x))$ is stable.

\section{Kink families in non-linear $(\mathbb{S}^1\times \mathbb{S}^1)$-Sigma models with different number of vacua}

\label{sec:3}

We are interested in investigating the kink variety in non-linear $(\mathbb{S}^1\times \mathbb{S}^1)$-Sigma models which involve a different number of vacua. In particular, we shall thoroughly explore in the next sections three different models whose potential terms $V(\theta,\varphi)$ have, respectively, eight, four and two vacuum points. In general, stability of non-topological kinks cannot be guaranteed by topological arguments because this type of solutions asymptotically begins and ends at the same vacuum. The local behavior of the potential can make these solutions stable against small fluctuations (linear stability). An example of this possibility can be found in reference \cite{Alonso2020}, where a deformation of the MSTB model comprises a linearly stable non-topological kink. The target space in this model was $\mathbb{R}^2$. However, these solutions may decay to the vacuum if a large enough fluctuation is applied in contrast to the topological partners, whose stability is protected by topological constraints. This type of conditions could be recovered for non-topological solutions when the potential term involves a singularity surrounded by the kink, see \cite{Alonso2007}, or when the target space is itself non-simply connected. The latter will be the present case, where the target space is the torus $\mathbb{S}^1\times \mathbb{S}^1$ and non-null homotopic non-topological kinks are found. Therefore, with the purpose of exploring the kink variety in non-linear $(\mathbb{S}^1\times \mathbb{S})$-Sigma models the following set of superpotentials is proposed
\begin{equation}
W_{n_1,n_2}(\theta,\varphi)=m (R+r \sin n_1 \theta) \sin (n_2\varphi) \, ,  \label{superpotential}
\end{equation}
where $m\in \mathbb{R}$, $n_1\in \mathbb{Z}$ and  $n_2$ is an integer or a half-integer. These conditions on $n_i$ are chosen to guarantee the physical meaning of the potentials (\ref{ps})
\begin{equation}
V_{n_1,n_2}(\theta,\varphi)= \frac{m^2}{2} \Big[ n_1^2 \cos^2(n_1\theta) \sin^2 (n_2\varphi) +\frac{n_2^2 \cos^2 (n_2\varphi) (R+r \sin(n_1\theta))^2}{(R+r \sin (\theta))^2} \Big] \label{potential}
\end{equation}
 associated to the superpotencials (\ref{superpotential}) defined on the torus. From (\ref{ps}) it is clear that the set of vacua ${\cal M}$ (zeroes of $V$) must comply with the conditions
\[
\frac{\partial W_{n_1,n_2}}{\partial \theta}= m n_1r \cos(n_1 \theta) \sin(n_2\varphi) =0 \, ,\hspace{0.5cm} \frac{\partial W_{n_1,n_2}}{\partial \varphi} = m n_2\cos (n_2\varphi) [R+r \sin (n_1\theta)]=0\, ,
\]
which lead to a total of $|4 n_1 n_2|$ vacuum points distributed on the torus
\[
{\cal M}_{n_1,n_2}= \Big\{ \Sigma_{k_1,k_2} = (\theta_{k_2}, \varphi_{k_1})= \Big(\frac{\pi}{2n_1}+\frac{k_1 \, \pi}{n_1} \, , \, \frac{\pi}{2n_2} +\frac{k_2 \, \pi}{n_2}  \Big) \quad | \quad k_i \in \mathbb{Z} \, \Big\}\, .
\]
Indeed, the number of vacuum points and kinks traveling between them in this set of superpotentials will depend on $n_1$ and $n_2$. On the other hand, the BPS kinks must comply with the first order differential equations (\ref{bpsb}), which in our case read
\begin{equation}
\frac{d \theta}{dx} = (-1)^a \, \frac{m \, n_1}{r} \cos (n_1\theta) \sin (n_2\varphi) \, ,\hspace{0.5cm} \frac{d \varphi}{dx} = (-1)^a \, m \, n_2 \frac{\cos (n_2\varphi) (R+r\sin (n_1\theta))}{(R+r\sin \theta)^2} \, . \label{eq:PrimerOrden}
\end{equation}
From (\ref{eq:PrimerOrden}) the kink orbit flow in the phase plane is determined by the equation
\begin{equation}
   \frac{d\theta}{d\varphi} = \frac{n_1 \cos (n_1\theta) (R+r\sin \theta)^2 \tan (n_2\varphi)}{n_2r(R+r\sin (n_1\theta))} \, . \label{eq:orbita}
\end{equation}
Despite the fact that (\ref{eq:orbita}) is a separable first order differential equation, the analytical identification of the whole kink orbit variety is not possible for general values of $n_1$ and $n_2$. This calculation is feasible for singular kinks for which one of its angular coordinates remains constant. As a consequence two different classes of singular kinks can be distinguished:
\begin{enumerate}
        \item $\Phi$-kinks: Loops traced on the torus with the $\theta$-variable fixed as $\theta=\frac{\pi}{2 n_1}+\frac{k_1}{n_1}$ cross each through one or more vacuum points. Each of these curves between vacua will correspond to a $\Phi$-kink. If we substitute the previously fixed value of $\theta$ into the first order equations (\ref{eq:PrimerOrden}) we obtain the expression for $\Phi_K^{(k_1,k_2)}(x) = (\theta_K(x),\varphi_K(x))$
    \begin{equation}
      \hspace{-0.38cm} \Phi_K^{(k_1,k_2)}(x) = \left( \frac{2k_1+1}{2 n_1} \pi \,\, , \,\, \frac{k_2+1}{n_2} \pi + \frac{1}{n_2}{\rm Gd} \left[(-1)^{a+k_2+1} \, n_2^2 \, A_{k_1} \bar{x})\right]\right)\, ,
        \label{eq:SingularPhi}
    \end{equation}
    where $\bar{x}=x-x_0$, ${\rm Gd}\left[y\right]=-\frac{\pi}{2} + 2 \arctan{e^y}$ denotes the Gudermannian function and
    $$A_{k_1}=\frac{m \,\left(R+(-1)^{k_1} r\right)}{(R+r \sin{\frac{2k_1+1}{2n_1}\pi})^2}\, .$$
    The integration constant $x_0$ can be understood as the kink center. In this case, kinks trace pieces of loops around the torus center, travelling between vacua located at $\varphi_K=\frac{2k_2+1}{2 n_2}$ and $\varphi_K=\frac{2k_2+3}{2 n_2}$. The total energy of the kinks (\ref{eq:SingularPhi}) is given by the relation
    \begin{equation}
    E[\Phi_K^{(k_1,k_2)}] =2 m \left[ R+(-1)^{k_1}\,r \,\right] \, . \label{ener1}
    \end{equation}
    The formula (\ref{ener1}) suggests that there are two different groups of $\Phi$-kinks determined by the value of the magnitude $k_1 \,{\rm mod}\, 2$ in the formula (\ref{eq:SingularPhi}), that is, depending on whether $k_1$ is even or odd. To distinguish these two kink subtypes we shall employ the notation $\Phi_K^{[0]}(x)$ and $\Phi_K^{[1]}(x)$ based on the modular arithmetic $[k_1]=k_1 \,{\rm mod}\, 2$. Note that the $\Phi_K^{[0]}(x)$-kinks are more energetic than the $\Phi_K^{[1]}(x)$-kinks, that is,
    \[
    E[\Phi_K^{[0]}]\,\,>\,\,E[\Phi_K^{[1]}] \, .
    \]

     \item $\Theta$-kinks: Vacuum points can also be connected by pieces of loops defined by the condition $\varphi=\frac{\pi}{2 n_2}+\frac{k_2}{n_2} \pi$. In this case the second equation in (\ref{eq:PrimerOrden}) holds trivially and the first one provides us with kink profiles $\Theta_K^{(k_1,k_2)}(x)= (\theta_K(x),\varphi_K(x))$ that travel between vacua located at $\theta_K=\frac{2k_1+1}{2 n_1}$ and $\theta_K=\frac{2k_1+3}{2 n_1}$
     \begin{equation}
     \hspace{-0.44cm}
	\Theta_K^{(k_1,k_2)}(x) = \left( \frac{k_1+1}{n_1}\pi +\frac{1}{n_1}{\rm Gd}\left[ \frac{(-1)^{a+k_1+k_2+1}\, m n_1^2}{r} \,\bar{x}\right],\frac{2k_2+1}{2n_2}\pi\right)\, ,
\label{eq:SingularTheta}
     \end{equation}
    whose total energy is $k_i-$independent
    \[
    E[\Theta_K]=2 m r \, .
    \]
	It can be checked that the energies of the previous singular kinks comply with the following sum rule
	\begin{equation}
	E[\Phi_K^{[0]}]\,\, = \,\,E[\Phi_K^{[1]}] \,\, + \,\, 2 E[\Theta_K] \, . \label{sumrule}
	\end{equation}
	
    \end{enumerate}

    While $\Phi_K^{[0]}$ is always the most energetic singular kink, the second most energetic one will be $\Phi_K^{[1]}$ or $\Theta_K$ depending on the radii of the torus $R$ and $r$. In particular, $E[\Phi_K^{[1]}] \, \geq \, E[\Theta_K]$ when $R\geq 2r$ and $E[\Phi_K^{[1]}] \, < \, E[\Theta_K]$ when $R<2r$. As previously mentioned, in order to describe more thoroughly the structure of the kink varieties in different non-linear $(\mathbb{S}^1\times \mathbb{S}^1)$-Sigma models with a distinct disposition of vacua on the torus, the following cases of the potential (\ref{potential}) shall be studied in detail:

    \begin{itemize}

        \item Case $1$: Values $n_1=1$ and $n_2=2$ are chosen to construct a potential with eight vacuum points. Sixty four disjoint topological sectors arise in the configuration space, but only twenty four will contain kinks. Sixteen singular kinks and antikinks are found traveling between vacuum points and eight families of kinks will appear between them. All of them will be topological and linearly stable.

    	\item Case $2$: Seeking to reduce the number the vacuum points of the model, the values $n_1=n_2=1$ are taken into the potential (\ref{potential}). Four vacua and sixteen topological sectors appear, eight of which are empty. Only eight singular topological kinks and antikinks emerge and only four families of topological kinks are found. Once again, all kinks are linearly stable.
    	
    	\item Case $3$: Lastly, the minimum number of vacua present in a model of the type (\ref{potential}) is sought while ensuring its physical meaning on the torus. This number is two and it can be found for example when $n_1=1$ and $n_2=\frac{1}{2}$. For this choice of $n_1$ and $n_2$, four topological sectors arise, all of them non-empty. Two topological singular kinks and antikinks are found but in this case so are two non-topological singular kinks and antikinks. Furthermore, two families of non-topological kinks will emerge. All of them are linearly stable. Indeed, this represents another example of not only linearly stable but globally stable non-topological kinks. This global stability as mentioned before is ensured by the non-simply connectedness of the chosen target manifold.
    \end{itemize}

\subsection{Kink variety for a non-linear $(\mathbb{S}^1\times \mathbb{S}^1)$-Sigma model with eight vacua}

\label{sec:3a}

\begin{figure}[ht]
\centerline{\includegraphics[height=4.5cm]{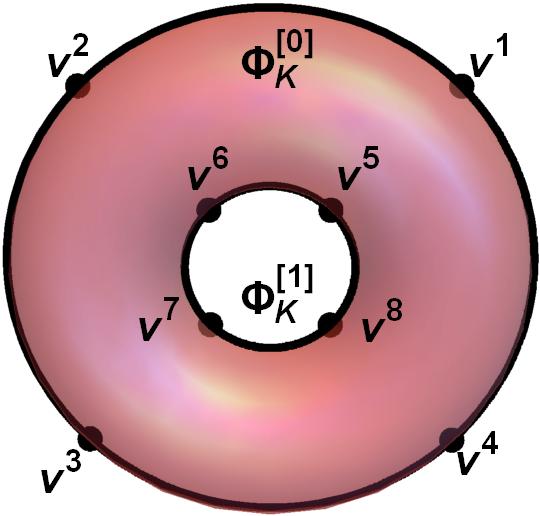} \hspace{0.5cm}
\includegraphics[height=4.5cm]{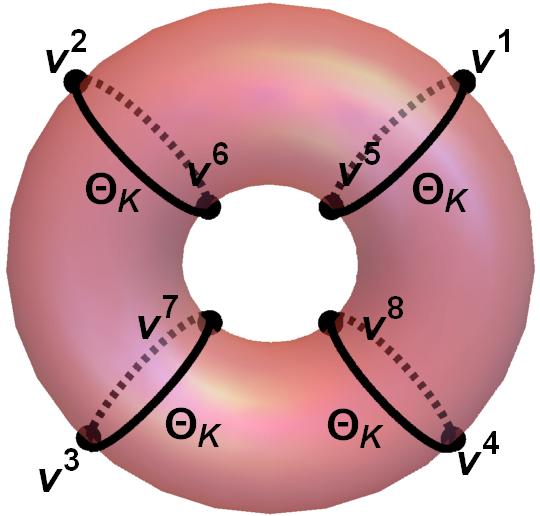}}
\caption{\small Orbits of $\Phi_K^{[0]}$ and $\Phi_K^{[1]}-$kinks (left) and those of $\Theta-$kinks (right). Dotted lines represent curves on the hidden side of the torus. Notice how $\Phi_K$ and $\Theta-$kink orbits split the torus into eight regions, which are equivalent by symmetry.}\label{figure:Z1y2}
\end{figure}

In this section the previously mentioned Case $1$ will be thoroughly explored. Recall that this model is determined by fixing $n_1=1$ and $n_2=2$ in the general framework introduced in the previous section. The superpotential is given by
\begin{equation}
W_{1,2}(\theta,\varphi)=m(R+r\sin \theta)\sin 2\varphi \, ,  \label{superpotentialA}
\end{equation}
whereas the potential term is expressed as
\begin{equation}
V_{1,2}(\theta,\varphi)= \frac{m^2}{2} \left[\cos^2\theta \sin^2 2\varphi + 4 \cos^2 2\varphi \right]  \, . \label{potentialZ}
\end{equation}
The set of vacua ${\cal M}$ contains eight points symmetrically separated on the torus

\[  {\cal M}_{1,2}=
      \textstyle \left\{v^1=\left(\frac{\pi}{2},\frac{\pi}{4}\right); v^2=\left(\frac{\pi}{2},\frac{3\pi}{4}\right) ; v^3=\left(\frac{\pi}{2},\frac{5\pi}{4}\right) ; v^4=\left(\frac{\pi}{2},\frac{7\pi}{4}\right);\right. \]
      \[\hspace{2.7cm}
      \left.v^5= \textstyle \left(\frac{3\pi}{2},\frac{\pi}{4}\right); v^6=\left(\frac{3\pi}{2},\frac{3\pi}{4}\right) ; v^7=\left(\frac{3\pi}{2},\frac{5\pi}{4}\right) ; v^8=\left(\frac{3\pi}{2},\frac{7\pi}{4}\right)\right\} \, , \\
 \]
whose distribution in the target space $\mathbb{S}^1\times\mathbb{S}^1$ can be seen in Figure \ref{figure:Z1y2}. The fluctuation operator associated to these constant solutions is given by
\[
{\cal H}[v^1,v^2,v^3,v^4] = -{\mathbb I} \frac{d^2}{dx^2} + \left( \begin{array}{cc}\frac{m^2}{r^2} & 0 \\ 0 & \frac{16 m^2}{(R + r)^2} \end{array} \right) \ ,\quad {\cal H}[v^5,v^6,v^7,v^8] = -{\mathbb I} \frac{d^2}{dx^2} + \left( \begin{array}{cc}  \frac{m^2}{r^2} & 0 \\ 0 & \frac{16 m^2}{(R - r)^2} \end{array} \right)\hspace{0.5cm},
\]
which clearly means that the members of $ {\cal M}_{1,2}$ are minima of the potential (\ref{potentialZ}). Notice that this potential involves certain angular symmetries. Indeed, translations $\theta \to \theta+\pi$ and $\varphi \to \varphi+\frac{\pi}{2}$ leave invariant
	$V_{1,2}[\theta,\varphi]$. These two discrete transformations generate the symmetry group of the action $G=\mathbb{Z}_2\otimes\mathbb{Z}_4$. This means that the torus is divided into eight regions where the structure of the kink variety is the same. Kinks defined in one of these regions can be identified with other kinks in other regions by applying the previously shown symmetries. In this model the kink variety includes eight singular topological kinks and antikinks. One family of topological kinks also emerges, which is replicated in all previously mentioned equivalent regions. Let us discuss in more detail these solutions:

	\noindent $\bullet$ $\Phi$-kinks: These singular kinks can be identified from the general expression (\ref{eq:SingularPhi}). In particular, these solutions will connect different vacua depending on $k_1=0,1$ and $k_2=0,1,2,3$
\begin{equation}
\Phi_K^{(k_1,k_2)} (x)= \left(\frac{2k_1+1}{2} \pi\,\, , \,\, \frac{k_2+1}{2} \pi+ \frac{1}{2} ~ {\rm Gd}\left[(-1)^{a+k_2+1}\frac{4 m \bar{x}}{R+(-1)^{k_1}r}\right] \, \right) \, .
\end{equation}
This means that $\Phi_K^{[0]}$-kinks and $\Phi_K^{[1]}$-kinks belong to different topological sectors, see Figure \ref{figure:Z1y2}. While $[k_1]$ distinguishes between the``outer and inner circumferences''on the torus, different values of $k_2$ mod $4$ discriminate between pieces of the circumference. Explicitly, these kinks connect vacua $v^{i}$ to $v^{[i+1]_4}$ and $v^{[i-1]_4}$ and vacua $v^{i+4}$ to $v^{[i+1]_4+4}$ and $v^{[i-1]_4+4}$ where $i=1,2,3,4$, and $[\cdot]_4$ stands for module $4$. The kink energy densities are all localized around a point describing a single lump, although their total energies are different. From the previous section, the following relations hold
\[
E[\Phi_K^{[0]}]= 2m(R+r) \, ,\hspace{1cm} E[\Phi_K^{[1]}]= 2m(R-r) \, .
\]

The study of the linear stability for the $\Phi$-kinks leads to the spectral problem
\begin{equation}
{\cal H}[\Phi_K^{[k_1]} (x)] \,\Psi_n (z) = \frac{\left(R+(-1)^{k_1} r\right)^2}{16 m^2} \, \omega_n^2 \, \Psi_n(z) \, , \hspace{1cm} k_1=0,1 \, , \label{hessAZ}
\end{equation}
where the non-null components of the second order small kink fluctuation ${\cal H} [\Phi_K^{[k_1]} (x)]$ read as
\begin{eqnarray*}
{\cal H}_{11} [\Phi_K^{[k_1]} (x)]&=& -\frac{d^2}{dz^2} + \frac{\left(R+(-1)^{k_1} r \right)^2}{16 r^2} \left(1-\frac{R+ (-1)^{k_1} 5 r}{R+ (-1)^{k_1} r}\,{\rm sech}^2 z\right) \, ,\\
{\cal H}_{22}[\Phi_K^{[k_1]} (x)] &=& -\frac{d^2}{dz^2} + \left(1-2 \,{\rm sech}^2 z\right)\, .
\end{eqnarray*}
Notice that these expressions read the same for any kink orbit with different $k_2$. This was expected given the present symmetries and the distribution of these $\Phi-$kinks. This spectral problem derives from the general form (\ref{spectralproblem}) with the space variable change
\[
z= \frac{4 m\,\overline{x}}{R+(-1)^{k_1} r} \, .
\]
Fortunately, the spectral problem (\ref{hessAZ}) reduces to two one-dimensional problems associated with Schr{\"o}dinger type differential operators with two different P\"osch-Teller potential wells. These problems are exactly solvable.

\begin{itemize}
    \item Spectrum of ${\cal H}_{22}[\Phi_K^{[k_1]} (x)]$: The discrete spectrum of the operator ${\cal H}_{22}[\Phi_K^{[k_1]} (x)]$ consists only of a zero mode (whose eigenvalue is zero) and the continuous spectrum emerges on the threshold value 1.
    \item Spectrum of ${\cal H}_{11}[\Phi_K^{[0]} (x)]$: The structure of the spectrum associated to the operator ${\cal H}_{11}[\Phi_K^{[k_1]} (x)]$ is more complex. For the case $k_1=0$, the operator ${\cal H}_{11}[\Phi_K^{[0]} (x)]$ has the discrete spectrum $\omega_n^2=\frac{n}{2}(\frac{R}{r}+1)-n^2$ for non-negative integers $n<\frac{1}{4} \left(\frac{R}{r}+1\right)$. Hence, $\lceil \frac{1}{4}\left(\frac{R}{r}+1\right) \rceil$ discrete states appear, where $\lceil x \rceil\equiv {\rm min} \left\{n\in \mathbb{Z} ~ | ~ n \geq x\right\}$ stands for $x$ if $x\in \mathbb{Z}$ or the immediate integer value above otherwise. Therefore, the number of discrete states grows as the ratio between the major and minor radius increases. The lowest of these modes is a zero mode, that is, the lowest eigenvalue of ${\cal H}_{11}[\Phi_K^{[0]} (x)]$ is zero. The continuous spectrum emerges on the threshold value $\frac{1}{16}(\frac{R}{r}+1)^2$.
    \item Spectrum of ${\cal H}_{11}[\Phi_K^{[1]} (x)]$: For the case $k_1=1$, the operator ${\cal H}_{11}[\Phi_K^{[1]} (x)]$ has a discrete spectrum $\omega_n^2=\frac{n+1}{2}(\frac{R}{r}-3-2n)$ for non-negative integers $n<\frac{1}{4}\left(\frac{R}{r}-5\right)$ and a continuous spectrum rising at the value $\frac{1}{16}(\frac{R}{r}-1)^2$. Notice that it only exhibits discrete spectrum if $\frac{R}{r}>5$, in particular $\lceil \frac{1}{4}\left(\frac{R}{r}-5\right) \rceil$ discrete states.
\end{itemize}
  In conclusion the operator ${\cal H}[\Phi_K^{[k_1]} (x)]$ has no negative eigenvalues and therefore $\Phi_K (x)$-kinks are linearly stable. In addition, the $\Phi_K^{[0]} (x)$ has two zero modes, one of them is associated with a translational symmetry but the other one suggests the existence of a kink family in the same topological sector. This possibility is investigated below.

\noindent $\bullet$ $\Theta$-kinks: Trying $\varphi$-constant orbits in the first order equations (\ref{eq:PrimerOrden}) of the current model leads us to four topological kinks together with their antikinks
\begin{equation}
\Theta_K^{(k_1,k_2)} (x) =  \left((k_1+1) \pi+ {\rm Gd}\left[(-1)^{a+k_1+k_2+1}\frac{m \bar{x}}{r}\right]\hspace{0.2cm} , \hspace{0.2cm} \frac{2 k_2+1}{4} \pi\right) \, ,
\end{equation}
where $ k_1=0,1$ and $k_2=0,1,2,3$. Kink orbits correspond to circular arcs associated to the minor radius. Given that two vacua are present in each circumference, eight topological kinks and anti-kinks will link ``exterior and interior'' vacuum points, see Figure \ref{figure:Z1y2}. While $k_1$ mod $2$ distinguishes between the visible and hidden side of the torus, $k_2$ mod $4$ differentiates between circunferences. Vacua $v^i$ will be doubly connected to $v^{[i+4]_8}$ with $i=1,\dots, 8$ and where $[\cdot]_8$ stands for module $8$. As it has been shown, the energy of all the $\Theta_K(x)$-kinks is identical:
\[
E[\Theta_K]=2m r \, .
\]
In this case, the analysis of the linear stability for the $\Theta$-kinks leads to the spectral problem
\begin{equation}
{\cal H}[\Theta_K (x)] \,\Psi_n (z) = \frac{r^2}{m^2} \, \omega_n^2 \, \Psi_n(z) \, , \label{hessA1}
\end{equation}
where the non-null components of the second order small kink fluctuation ${\cal H} [\Theta_K (x)]$ read as
\begin{eqnarray*}
	{\cal H}_{11} [\Theta_K (x)]&=& -\frac{d^2}{dz^2} + \left(1-2 ~ {\rm sech}^2 z\right) \, ,\\
	{\cal H}_{22}[\Theta_K (x)] &=& -\frac{d^2}{dz^2} -\frac{2 r ~
   {\rm sech}^2 z}{R+r \tanh z} \frac{d }{dz} + \frac{4r^2
   \left(4-{\rm sech}^2 z\right)}
   {\left(R+ r \tanh
   z\right)^2}
\end{eqnarray*}
and where the variable change $z=(-1)^{a+k_2}\frac{m x}{r}$ has been employed. Again, the spectral problem (\ref{hessA1}) reduces to two independent one-dimensional problems associated with the diagonal components of ${\cal H}[\Theta_K (x)]$. The first component ${\cal H}_{11} [\Theta_K (x)]$ involves the presence of a zero mode and a continuous spectrum arising at the threshold value $1$. On the other hand the analysis of spectral problem for the second component ${\cal H}_{22} [\Theta_K (x)]$ is again far more complicated. Since it is not analytically solvable, numerical methods have been used to extract the spectral information of this operator. No negative eigenvalues of $\omega^2$ are found and therefore $\Theta_K(x)-$kinks are linearly stable.

\begin{figure}[h]
\centerline{\includegraphics[height=4.8cm]{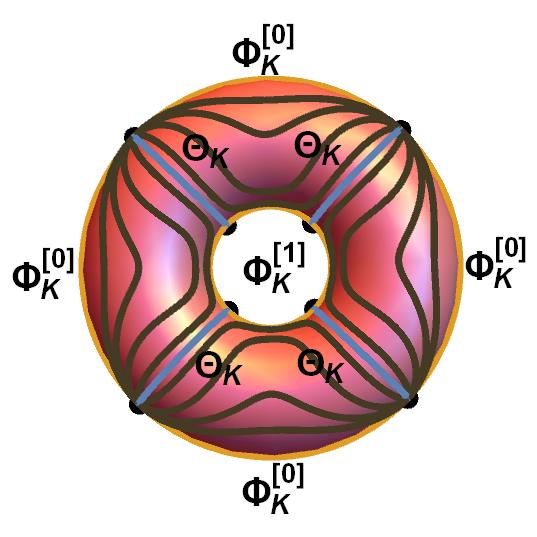}  \hspace{0.5cm} \includegraphics[width=6cm,height=4.5cm]{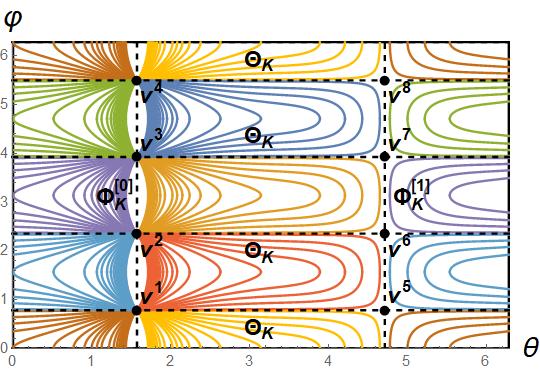}}
\caption{\small Some orbits of the family $\Sigma_K^{(k_1,k_2)} (x;\gamma)$ have been depicted on the torus and on the $\theta-\varphi-$plane. Orbits on the hidden side have been omitted in the first picture. Note in both representations how singular kinks are recovered as limits of this family according to (\ref{familylimit}).}\label{Familia8}
\end{figure}

\noindent $\bullet$ Kink families: As previously mentioned, the fact that the $\Phi_K^{[0]}(x)$-kinks have a second zero mode suggests that these kinks are limit members of a family of kinks. Members of this family shall be denoted as $\Sigma_K^{(k_1,k_2)} (x;\gamma)$, where $\gamma$ is employed to label every kink member. In this case, the kink orbits can be obtained by solving the orbit flow equation
\begin{equation}
\frac{d\theta}{d\varphi} = \frac{\cos \theta (R+r\sin \theta) \tan (2\varphi)}{2 r} \, , \label{eq:orbitaA}
\end{equation}
derived from the general equation (\ref{eq:orbita}). Their analytical expression is given by
\begin{equation}
\left|\cos 2\varphi\right|=   \frac{1}{\gamma} \frac{\Big| 1-\sin \theta \Big|^{\frac{2r}{(R+r)}} \cdot \Big| R+r \sin\theta \Big|^{\frac{4r^2}{R^2-r^2}} }{|1+\sin \theta|^{\frac{2r}{(R-r)}}} \, ,
\label{eq:orbA}
\end{equation}
where the parameter $\gamma\in (0,\infty )$ is constant. All these kinks connect ``outer'' vacua, $v^i$ to $v^{[i+1]_4}$ and $v^{[i-1]_4}$ where $i=1,2,3,4$. The energy of all members of these families is the same $E\left[\Sigma_K\right]=2m(R+r)$. In Figure \ref{Familia8} several kinks belonging to this family have been depicted on the torus and on the $\theta-\varphi-$plane.

 Singular kinks are recovered as certain limits of this family. The limit $\gamma\rightarrow 0$ demands $\theta=\frac{\pi}{2}$, letting $\varphi$ vary freely between the outer vacua, which precisely defines the $\Phi_K^{[0]}(x)$-kink orbits. On the other hand, when $\gamma\to\infty$ there are two possible solutions of (\ref{eq:orbA}). One of the possibilities is that $\varphi$ takes one of the values $\varphi=\frac{\pi}{4}, \frac{3\pi}{4}, \frac{5\pi}{4}, \frac{7\pi}{4}$ whereas the variable $\theta$ is free to vary with $x$. This description leads to the $\Theta_K (x)$-kinks. The second possibility corresponds to the condition $\theta=\frac{3\pi}{2}$, which annihilates the denominator of (\ref{eq:orbA}). This takes us to the indeterminate form $\frac{\infty}{\infty}$ and $\varphi$ can vary with $x$. This is the $\Phi_K^{[1]}(x)$-kink orbit. Symbolically, we can write
\begin{equation}
\lim_{\gamma \rightarrow 0} \Sigma_K^{(k_1,k_2)} (x;\gamma) \equiv \Phi_K^{[0]}(x) \, ,\hspace{0.6cm} \lim_{\gamma \rightarrow \infty} \Sigma_K^{(k_1,k_2)} (x;\gamma) \equiv \Theta_K (x) \cup \Phi_K^{[1]}(x) \cup \Theta_K (x) \, . \label{familylimit}
\end{equation}
These limits comply with the previously found energy sum rule for singular kinks (\ref{sumrule}), that is, $E[\Sigma_K]=E[\Phi_K^{[0]}]\,\, = \,\,E[\Phi_K^{[1]}] \,\, + \,\, 2 E[\Theta_K]$.

In sum, a family of topological kinks is replicated on the eight regions of the torus. The limits of this family coincide with the singular kinks, which as mentioned before delimit these eight regions. Since there is no intermediate points in these kinks where the flow is undefined, there are no other conjugate points where the kink orbits intersect. This fact ensures according to the Morse index theorem, see \cite{Gonzalez}, that all these kinks are stable.

\subsection{Kink variety for a non-linear $(\mathbb{S}^1\times \mathbb{S}^1)$-Sigma model with four vacua}

\label{sec:3b}

\begin{figure}[ht]
\centerline{\includegraphics[height=4.5cm]{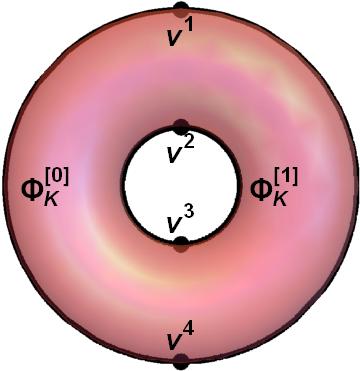} \hspace{0.5cm}
\includegraphics[height=4.5cm]{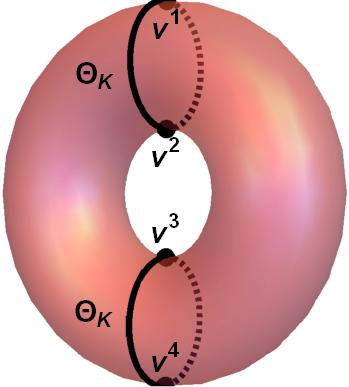}}
\caption{\small Orbits of $\Phi_K^{[0]}$ and $\Phi_K^{[1]}-$kinks (left) and those of $\Theta-$kinks (right). Dotted lines represent curves on the hidden side of the torus. Notice how $\Phi_K$ and $\Theta-$kink orbits split the torus into four regions, which are equivalent by symmetry.}\label{figure:A1y2}
\end{figure}

In this section Case $2$ will be explored, where $n_1=n_2=1$ are chosen to fix the number of vacua of the model to four. This choice of $n_i$ leads us to a superpotential which is $2 \pi-$periodic in both angles
\begin{equation}
W_{1,1}(\theta,\varphi)=m(R+r\sin \theta)\sin \varphi \, ,  \label{superpotentialA2}
\end{equation}
whereas the potential term is $\pi-$periodic and is expressed as
\begin{equation}
V_{1,1}(\theta,\varphi)= \frac{m^2}{2} \left[\cos^2\theta \sin^2 \varphi + \cos^2 \varphi \right]  \, . \label{potentialA2}
\end{equation}
Indeed, the set of vacua ${\cal M}$ consists now of only four points
\begin{eqnarray*}
{\cal M}_{1,1} & = & \textstyle \{ v^1=(\frac{\pi}{2},\frac{\pi}{2}); v^2=(\frac{3\pi}{2},\frac{\pi}{2}) ; v^3=(\frac{3\pi}{2},\frac{3\pi}{2}) ; v^4=(\frac{\pi}{2},\frac{3\pi}{2})\} \, ,
\end{eqnarray*}
whose distribution in the target space $\mathbb{S}^1\times\mathbb{S}^1$ can be seen in Figure \ref{figure:A1y2}. As before, these points are minima of the potential (\ref{potentialA2}) because the fluctuation operator associated to this points is given by
\[
{\cal H}[v^1,v^4] = -{\mathbb I} \frac{d^2}{dx^2} + \left( \begin{array}{cc} \frac{m^2}{r^2} & 0 \\ 0 & \frac{m^2}{(R + r)^2} \end{array} \right) \hspace{0.5cm},\hspace{0.5cm} {\cal H}[v^2,v^3] = -{\mathbb I} \frac{d^2}{dx^2} + \left( \begin{array}{cc} \frac{m^2}{r^2} & 0 \\ 0 & \frac{m^2}{(R - r)^2} \end{array} \right)\hspace{0.5cm}.
\]
As expected, reducing the number of vacua also affects the symmetries of the new potential (\ref{potentialA2}). $\pi$-translations in now both angles, $\theta \to \theta+\pi$ and $\varphi \to \varphi+\pi$, leave invariant
	$V_{1,1}[\theta,\varphi]$. These generate the symmetry group of the action $G=\mathbb{Z}_2\otimes\mathbb{Z}_2$. This means that in this case the torus is divided into only four identical regions with regard to the kink variety. The kink variety in this model includes eight singular topological kinks and antikinks. Moreover, a family of topological kinks will be found in each of the four regions. Let us examine these solutions:

\noindent $\bullet$ $\Phi$-kinks: In this case singular kinks will connect $v^1$ to $v^4$ or $v^3$ to $v^2$ depending on $k_1=0,1$
\begin{equation}
\Phi_K^{k_1} (x)= \left(\frac{2k_1+1}{2} \,\, , \,\, (k_2+1) \pi+(-1)^{a+k_2+1}{\rm Gd}\left[\frac{m \bar{x}}{R+(-1)^{k_1}r}\right] \, \right) \, ,
\end{equation}
and where  $k_2=0,1$. This means that $\Phi_K^{[0]}$-kinks and $\Phi_K^{[1]}$-kinks also belong to different topological sectors in this case, see Figure \ref{figure:A1y2}. Note that while $k_1$ discriminate between the outer and inner arcs, $k_2$ distinguishes between the two pieces of the circumferences. The kink energy densities are localized describing a single lump, with energies
\[
E[\Phi_K^{[0]}]= 2m(R+r) \, ,\hspace{1cm} E[\Phi_K^{[1]}]= 2m(R-r) \, .
\]

The linear stability of the $\Phi$-kinks is encoded in the spectral problem
\begin{equation}
{\cal H}[\Phi_K^{[k_1]} (x)] \,\Psi_n (z) = \frac{[R+(-1)^{k_1} r]^2}{m^2} \, \omega_n^2 \, \Psi_n(z) \, , \hspace{1cm} k_1=0,1 \, , \label{hessA0}
\end{equation}
where the non-null components of this operator ${\cal H} [\Phi_K^{[k_1]} (x)]$ have the form
\begin{eqnarray*}
{\cal H}_{11} [\Phi_K^{[k_1]} (x)]&=& -\frac{d^2}{dz^2} + \frac{(R+(-1)^{k_1} r)^2}{r^2} - \frac{(R+(-1)^{k_1}2r)(R+(-1)^{k_1} r)}{r^2} \, {\rm sech}^2 z  \, ,\\
{\cal H}_{22}[\Phi_K^{[k_1]} (x)] &=& -\frac{d^2}{dz^2} + 1-2 \,{\rm sech}^2 z \, .
\end{eqnarray*}
Similarly to Case $1$, this spectral problem matches that of the general form (\ref{spectralproblem}) when the following space variable change is used
\[
z= \frac{m\,\overline{x}}{R+(-1)^{k_1} r} \, .
\]
\begin{itemize}
    \item Spectrum of ${\cal H}_{22}[\Phi_K^{[k_1]} (x)]$: Exactly as in the previous case with eight vacua, the discrete spectrum of the operator ${\cal H}_{22}[\Phi_K^{[k_1]} (x)]$ contains exclusively a zero mode and the continuous spectrum threshold appears at $1$.

    \item Spectrum of ${\cal H}_{11}[\Phi_K^{[0]} (x)]$: Similarly, two cases for ${\cal H}_{11}[\Phi_K^{[k]} (x)]$ are considered depending at the value of $\left[k_1\right]$. For the case $k_1=0$, $\lceil(1+\frac{R}{r})\rceil$ discrete states are found, since the operator ${\cal H}_{11}[\Phi_K^{[0]} (x)]$ has a discrete spectrum $\omega_n^2=2n\left(\frac{R}{r}+1\right) -n^2$ for non-negative integers $n<\frac{R}{r}+1$. Again, amongst these states a zero mode is found. The continuous spectrum threshold is now at the value $(1+\frac{R}{r})^2$.

    \item Spectrum of ${\cal H}_{11}[\Phi_K^{[1]} (x)]$: For the case $k_1=1$, the discrete spectrum $\omega_n^2=(n+1)(\frac{2 R}{r}-3-n)$ emerges only when $\frac{R}{r}>2$ and for non-negative integers $n<\frac{R}{r}+2$. $\lceil \frac{R}{r}-2\rceil$ states arise until the continuous spectrum at the value $(\frac{R}{r}-1)^2$ is reached. Notice that the number of discrete states is again proportional to the ratio $\frac{R}{r}$ in all cases.
\end{itemize}

    Once more, no negative value of $\omega_n^2$ is found and therefore these kinks are linearly stable. The second zero mode suggests again the existence of a family of kinks, possibility that will be again explored below.

\noindent $\bullet$ $\Theta$-kinks: Four topological kinks together with their antikinks are found when the $\varphi$-constant condition is imposed in the first order equations (\ref{eq:PrimerOrden})
\begin{equation}
\Theta_K^{(k_1,k_2)} (x) =  \left((k_1+1) \pi+(-1)^{a+k_2+1} {\rm Gd}\left[\frac{m \bar{x}}{r}\right]\, , \hspace{0.4cm} \frac{\pi}{2} + k_2 \pi \right) \hspace{0.3cm},\hspace{0.3cm} \quad k_1,k_2=0,1 \, .
\end{equation}
These kink orbits correspond to circular arcs associated to the minor radius, see Figure \ref{figure:A1y2}.
These topological kinks and anti-kinks will link $v^1$ to $v^2$ and $v^3$ to $v^4$. Note that the $\Theta^{(k_1,0)}$-kinks asymptotically connect the vacua $v^1$ and $v^2$ whereas the $\Theta^{(k_1,1)}$-kinks join the points $v^3$ and $v^4$. On the other hand, the $\Theta^{(0,0)}$-kinks and $\Theta^{(1,0)}$-kinks are related by the Cartesian coordinate transformation $\phi_2 \leftrightarrow -\phi_2$. In sum, label $k_2$ distinguishes between the upper and lower circumference while $k_1$ differentiates between the kinks on one side and the kinks on the opposite side of the torus. Once again, the energy of all the $\Theta_K(x)$-kinks is identical:
\[
E[\Theta_K]=2m r \, .
\]
The second order small kink fluctuation for the $\Theta$-kinks leads to the spectral problem
\begin{equation}
{\cal H}[\Theta_K (x)] \,\Psi_n (z) = \frac{r^2}{m^2} \, \omega_n^2 \, \Psi_n(z) \, . \label{hessA2}
\end{equation}
The non-null components that must be analyzed to determine the linear stability of these kinks are again those of the diagonal
\begin{eqnarray*}
	{\cal H}_{11} [\Theta_K (x)]&=& -\frac{d^2}{dz^2} + 1-2 \,{\rm sech}^2 z \, ,\\
	{\cal H}_{22}[\Theta_K (x)] &=& -\frac{d^2}{dz^2} -\frac{2\, r \, {\rm sech}\, z}{R \cosh z + r \sinh z} \frac{d }{dz} + \frac{r^2 \sinh^2 z}{(R \cosh z + r\sinh z)^2} \, ,
\end{eqnarray*}
where the variable change $z=\frac{m x}{r}$ has been performed in the original spectral problem (\ref{spectralproblem}). Again, the spectral problem (\ref{hessA2}) reduces to two independent one-dimensional problems associated with the diagonal components of ${\cal H}[\Theta_K (x)]$. The first component ${\cal H}_{11} [\Theta_K (x)]$ involves the presence of a zero mode and a continuous spectrum arising at the threshold value 1. Exactly like the previous case with eight vacua, the analysis of the spectral problem for ${\cal H}_{22} [\Theta_K (x)]$ will be numeric since it is not analytically solvable. Numerical methods reveal the absence of negative eigenvalues of $\omega^2$, which implies the linear stability of these kinks.

\begin{figure}[ht]
\centerline{\includegraphics[height=4.5cm]{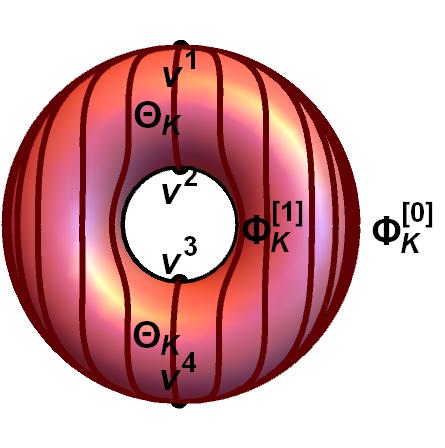}  \hspace{0.5cm} \includegraphics[width=6cm,height=4.5cm]{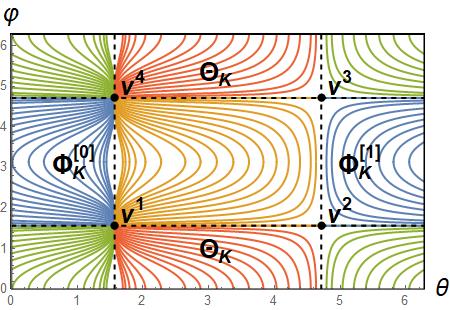}}
\caption{\small Some orbits of the family $\Sigma_K^{(k_1,k_2)} (x;\gamma)$ have been depicted on the torus and on the $\theta-\varphi-$plane. Orbits on the hidden side have been omitted in the first picture. Note in both representations how singular kinks are recovered as limits of this family according to (\ref{familylimit2}).}\label{Familia4}
\end{figure}

\noindent $\bullet$ Kink families: As in the previous section, the existence of a second zero mode for $\Phi_K^{[0]}(x)$-kinks prompts us to search for families of kinks in the four previously mentioned equivalent regions of the torus. These families of kink orbits $\Sigma_K^{(k_1,k_2)} (x;\gamma)$ are obtained by solving the orbit flow equation
\begin{equation}
\frac{d\theta}{d\varphi} = \frac{\cos \theta (R+r\sin \theta) \tan (\varphi)}{r} \, , \label{eq:orbitaB}
\end{equation}
derived again from the general equation (\ref{eq:orbita}). Their analytical expression is
\begin{equation}
\left|\cos \varphi\right|=  \Big[ \frac{1}{\gamma} \frac{\Big| 1-\sin \theta \Big|^{\frac{r}{2(R+r)}} \cdot \Big| R+r \sin\theta \Big|^{\frac{r^2}{R^2-r^2}} }{|1+\sin \theta|^{\frac{r}{2(R-r)}}}\Big] \, ,
\end{equation}
where $\gamma\in (0,\infty )$ is constant. All these kinks asymptotically connect the vacua $v^1$ and $v^4$ as shown in Figure \ref{Familia4}. The energy of all members of these families is the same $E\left[\Sigma_K\right]=2m(R+r)$. The asymptotic analysis of the kink orbits carried out in the previous section is valid in this case almost unaltered. The limit $\gamma\rightarrow 0$ corresponds to $\Phi_K^{[0]}(x)$-kink orbits and $\gamma\to\infty$ to a combination of $\Theta_K (x)$-kinks and $\Phi_K^{[1]}(x)$-kinks. These limits coincide formally with those of the previous case. However, the distribution and number of kinks traveling between vacua is different. In that sense, the same symbolic limits hold

\begin{equation}
\lim_{\gamma \rightarrow 0} \Sigma_K^{(k_1,k_2)} (x;\gamma) \equiv \Phi_K^{[0]}(x) \, ,\hspace{0.6cm} \lim_{\gamma \rightarrow \infty} \Sigma_K^{(k_1,k_2)} (x;\gamma) \equiv \Theta_K (x) \cup \Phi_K^{[1]}(x) \cup \Theta_K (x) \, , \label{familylimit2}
\end{equation}

\noindent which is again consistent with the energy sum rule $E[\Sigma_K]=E[\Phi_K^{[0]}]\,\, = \,\,E[\Phi_K^{[1]}] \,\, + \,\, 2 E[\Theta_K]$. Same analysis as in the previous model reveals the stability of these kinks due to the absence of intermediate conjugate points along these curves\footnote{It is interesting to remark that superpotential (\ref{superpotentialA2}) is nothing but the height function of the torus: $W_{1,1}(\theta,\varphi)= m\, \phi^3 (\theta,\phi)$. Thus kink orbits in this model can be interpreted as the flow lines of the height function, and, consequently, their stability is dictated by the corresponding Morse Theory. For the other two models studied in this paper, this kind of identification is also possible but considerably more complicated.}.

\subsection{Kink variety for a non-linear $(\mathbb{S}^1\times \mathbb{S}^1)$-Sigma model with two vacua}

\label{sec:3c}

\begin{figure}[ht]
\begin{center}
    \includegraphics[height=4.5cm]{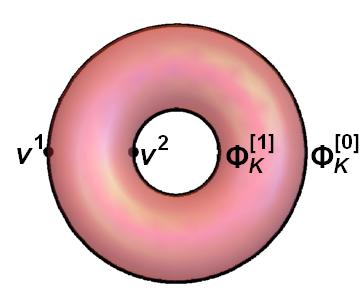} \hspace{0.5cm}
    \includegraphics[height=4.2cm]{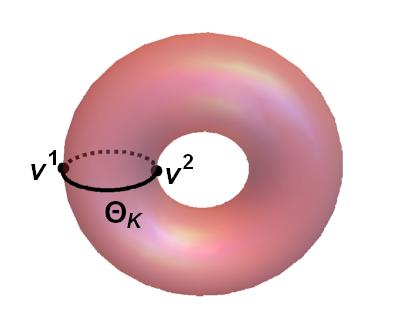}
    \caption{\small Orbits of $\Phi_K^{[0]}$ and $\Phi_K^{[1]}-$kinks (left) and those of $\Theta-$kinks (right). Dotted lines represent curves on the hidden side of the torus. Notice how $\Phi_K$ and $\Theta-$kink orbits split the torus into two regions, which are equivalent by symmetry.}\label{fig:ModBsing}
    \end{center}
\end{figure}

In Case $2$ the number of vacua had been reduced, but no non-topological kinks were found. Circumferences around the torus passing through vacua were comprised of two singular kinks. One way of searching for non-topological kinks in this family of potentials is making sure that some of these circumferences contain only one vacuum point. This can be achieved by an adequate choice of some $n_1$ and $n_2$. Indeed, the number of vacua $|4 n_1 n_2|$ is minimized when $|n_1|=1$ and $|n_2|=\frac{1}{2}$, and is equal to two. This is precisely the aim of Case $3$, where the kink variety for models whose potential is given by the general expression (\ref{potential}) with $n_1=1$ and $n_2=\frac{1}{2}$ shall be studied. This choice of $n_i$ produces a potential which is periodic on the torus
\begin{equation}
V_{1,\frac{1}{2}}(\theta,\varphi)= \frac{m^2}{2} \Big[\frac{1}{4} \cos^2 \left(\frac{\varphi}{2}\right)+ \cos^2 \theta \sin^2 \left(\frac{\varphi}{2}\right) \Big] \, , \label{potentialB}
\end{equation}
but that can be derived from a superpotential
\begin{equation}
W_{1,\frac{1}{2}}(\theta,\varphi)=m (R+r \sin  \theta) \sin \left(\frac{\varphi}{2} \right) \quad   \label{superpotentialB}
\end{equation}
 which does not comply with this property. This fact is precisely what allows the existence of BPS non-topological kinks. The new set ${\cal M}$ has only two vacuum points
\[
{\cal M}_{1,\frac{1}{2}}=\left\{ v^1= \left(\frac{\pi}{2},\pi \right)\,\, , \,\, v^2= \left(\frac{3\pi}{2},\pi \right) \right\} \, ,
\]
as it can be observed in Figure \ref{fig:ModBsing}. Now, the vacuum fluctuation operator is
\[
{\cal H}[v^1] = -{\mathbb I} \frac{d^2}{dx^2} + \left( \begin{array}{cc} \frac{m^2}{r^2} & 0 \\ 0 & \frac{m^2}{16(R + r)^2} \end{array} \right) \hspace{0.5cm},\hspace{0.5cm} {\cal H}[v^2] = -{\mathbb I} \frac{d^2}{dx^2} + \left( \begin{array}{cc} \frac{m^2}{r^2} & 0 \\ 0 & \frac{m^2}{16(R - r)^2} \end{array} \right)\hspace{0.5cm}.
\]
By reducing the number of vacua one of the symmetries present in previous cases has been removed. Only the angular symmetry $\theta \to \theta+\pi$ remains, which generates the symmetry group of the action $G=\mathbb{Z}_2$. This implies that two equivalent regions where the structure of the kink variety is the same will emerge. More specifically, such kink variety will include two singular non-topological kinks, two singular topological kinks and a family of non-topological kinks. A thorough description of these kinks is as follows:

\noindent $\bullet$ $\Phi$-kinks: Finally non-topological singular kinks are found from the general expression (\ref{eq:SingularPhi}). In fact, two types of non-topological kinks appear depending on the value $[k_1]$
\begin{equation}
\Phi_K^{[k_1]}(x)= \left( \frac{\pi}{2}+k_1 \pi \,\, , \,\, 2 \pi+ (-1)^{a+1} ~  2 ~ {\rm Gd}\left[\frac{m \overline{x}}{(R+(-1)^{k_1} r)}\right]\right) \, .
\end{equation}
Indeed, this expression characterizes singular non-topological kinks and their antikinks, which asymptotically connect the vacua $v^1$ and $v^2$ with themselves, see Figure \ref{fig:ModBsing}. No splitting of these circumferences into more than one solution is possible in this case since $2(k_2+1)$ can only take one value in $[0,2\pi)$. Notice that these kinks have been progressively absorbing adjacent kinks in all three cases until only one remained. Naturally, only one can emerge now in each circumference since only one vacuum point is present. The discussion of the energy is no different from the previous cases, the total energy of these solutions is given by
\[
E[\Phi_K^{[k_1]}]=2m(R+(-1)^{k_1} ~ r) \, .
\]
The analysis of the linear stability for the $\Phi_K^{[k_1]}(x)$-kinks for this case is analogous to previous $\Phi_K^{[k_1]}(x)$-kinks. The spectral problem reads almost identical
\begin{equation}
{\cal H}[\Phi_K^{[k_1]} (x)] \,\Psi_n (z) = \frac{16 [R+(-1)^{k_1} ~ r]^2}{m^2} \, \omega_n^2 \, \Psi_n(z) \, , \hspace{1cm} k_1=0,1 \, , \label{hessB0}
\end{equation}
where once again only diagonal components of the operator ${\cal H}[\Phi_K^{[k_1]} (x)]$ do not vanish
\begin{eqnarray*}
{\cal H}_{11}[\Phi_K^{[k_1]} (x)] &=& -\frac{d^2}{d z^2}+\frac{16(R+(-1)^{k_1} ~r)^2}{r^2}  \Big[ 1 -\frac{(4R+(-1)^{k_1} ~5r)}{4(R+(-1)^{k_1} ~r)}{\rm sech}^2 z\Big] \, ,\\
{\cal H}_{22}[\Phi_K^{[0]} (x)] &=&-\frac{d^2}{d z^2}+\left(1-2  ~ {\rm sech}^2 z\right)
\end{eqnarray*}
and the change of variable $z=\frac{m \bar{x}}{4(R+(-1)^{k_1} ~r)}$ has been included in (\ref{spectralproblem}).

\begin{itemize}
    \item Spectrum of ${\cal H}_{22}[\Phi_K^{[k_1]} (x)]$: As before, a zero mode is all the discrete spectrum of ${\cal H}_{22}[\Phi_K^{[k_1]} (x)]$ consists of and the continuous spectrum starts at $1$.

    \item Spectrum of ${\cal H}_{11}[\Phi_K^{[0]} (x)]$: On one hand, the operator ${\cal H}_{11}[\Phi_K^{[0]} (x)]$ has now a discrete spectrum $\omega_n^2=8 n \left(\frac{R}{r}+1\right) -n^2$ for non-negative integers $n<4\left(\frac{R}{r}+1\right)$. $\lceil4\left(\frac{R}{r}+1\right)\rceil$ discrete states appear before the continuous spectrum emerges at the threshold value $16\left(1+\frac{R}{r}\right)^2$.

    \item Spectrum of ${\cal H}_{11}[\Phi_K^{[1]} (x)]$: On the other hand, for ${\cal H}_{11}[\Phi_K^{[1]} (x)]$ the discrete spectrum $\omega_n^2=(n+1)\left(\frac{8R}{r}-9+n\right)$ is found only when $\frac{R}{r}>\frac{5}{4}$ and for non-negative integers $n<\frac{4R}{r}-5$. $\lceil \frac{4R}{r}-5\rceil$ discrete states are found before the continuous threshold appears at $16(\frac{R}{r}-5)^2$ in this case.
\end{itemize}
    Same argument as in the previous cases guarantees the linear stability of these kinks.

\noindent $\bullet$ $\Theta$-kinks: The remaining circumference that crosses vacua in this model contains two vacua. This corresponds to singular solutions (\ref{eq:SingularTheta})
\[
\Theta_K^{(k_1)} (x)= \left( (k_1+1) ~ \pi + (-1)^{a+k_1+k_2+1} {\rm Gd}\left[ \frac{m\overline{x}}{2r}\right]  \,\, , \,\,  \pi \right) \, ,
\]
which describe a couple of topological kinks and their antikinks asymptotically connecting the two vacua $v^1$ and $v^2$. The two values of $[k_1]$ give rise to two different orbits, one on the front side and one on the hidden side of the torus, see Figure \ref{fig:ModBsing}. As stated before, the total energy of these kinks is
\[
E[\Theta_K]=2 m r\, .
\]
Similarly, the spectral problem to analyze in this case has the form
\begin{equation}
{\cal H}[\Theta_K (x)] \,\Psi_n (z) = \frac{r^2}{m^2} \, \omega_n^2 \, \Psi_n(z) \, , \label{hessB3}
\end{equation}
where the non-null components of the operator ${\cal H}[\Theta_K (x)]$ are diagonal
\begin{eqnarray*}
	{\cal H}_{11}[\Theta_K (x)] &=&-\frac{d^2}{d z^2}+\left(1-2 \, {\rm sech}^2 z\right) \, , \\
	{\cal H}_{22}[\Theta_K (x)] &=&-\frac{d^2}{d z^2}- \frac{2 r \,{\rm sech}^2 z}{R+r\tanh z} \frac{d}{dz} +  \frac{r^2(1-4 \, {\rm sech}^2 z )}{16(R+r \tanh z)^2}
\end{eqnarray*}
and the change of variable $z=\frac{m \bar{x}}{r}$ has been made in (\ref{spectralproblem}). Again, for ${\cal H}_{11} [\Theta_K (x)]$ a zero mode is found and the continuous spectrum emerges at the threshold value $1$. Numerical analysis is once again necessary for the ${\cal H}_{22}[\Theta_K (x)]$ operator, for which no negative eigenvalue of $\omega^2$ is found. This ensures the linear stability of $\Theta_K (x)-$kinks.

\begin{figure}[ht]
\centerline{\includegraphics[height=4.5cm]{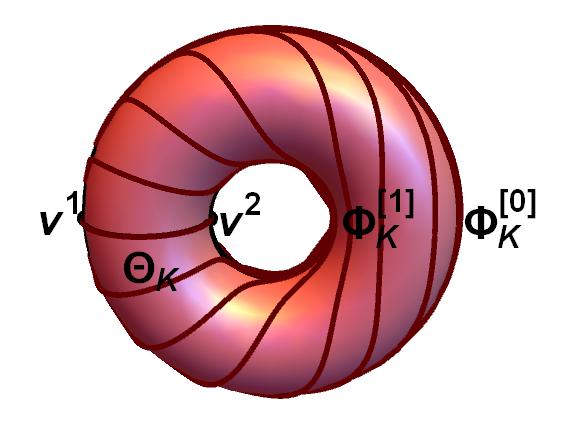}  \hspace{0.5cm} \includegraphics[width=6cm,height=4.5cm]{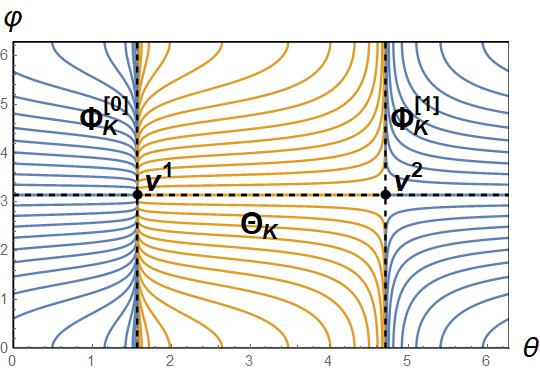}}
\caption{\small Some orbits of the family $\Sigma_K^{(k_1,k_2)} (x;\gamma)$ have been depicted on the torus and on the $\theta-\varphi-$plane. Orbits on the hidden side have been omitted in the first picture. Note in both representations how singular kinks are recovered as limits of this family.}\label{fig:fam2v}
\end{figure}

\noindent $\bullet$ Kink families: In Case $3$ the orbit flow equation (\ref{eq:orbita}) is again almost identical to those of the previous cases
\begin{equation}
\frac{d\theta}{d\varphi} = \frac{2\, \cos \theta \, (R+r\sin \theta)^2 \tan (\frac{\varphi}{2})}{r(R+r\sin \theta)} \label{eq:orbitaC}
\end{equation}
and can be integrated as well. The orbits for this kink family read now
\begin{equation}
\left|\cos \frac{\varphi}{2}\right|=\left[\frac{1}{\gamma} \frac{\Big| 1-\sin \theta \Big|^{\frac{r}{8(R+r)}} \cdot \Big| R+r \sin\theta \Big|^{\frac{r^2}{4(R^2-r^2)}} }{|1+\sin \theta|^{\frac{r}{8(R-r)}}}\right] \, ,
\end{equation}
where $\gamma\in \left(0,\infty\right)$ is again a constant that characterizes different kink orbits. All these kinks depart from $v^1$ and return to the same vacuum point, see Figure \ref{fig:fam2v}. Therefore, these two families, one on each side of the torus, are comprised of non-topological kinks. The energy of all members of these families is the same $E\left[\Sigma_K\right]=2m(R+r)$. When the parameter limit $\gamma\rightarrow0$ is considered then the singular $\Phi_K^{[0]}(x)$-kink is recovered whereas the limit $\gamma\rightarrow\infty$ leads to the concatenation of three kinks following the general form $\Theta_K (x) \cup \Phi_K^{[1]}(x) \cup \Theta_K (x)$. These are the same symbolic limits (\ref{familylimit}) found in previous sections, which obviously are again compatible with the energy sum rule $E[\Sigma_K]=E[\Phi_K^{[0]}]\,\, = \,\,E[\Phi_K^{[1]}] \,\, + \,\, 2 E[\Theta_K]$. Finally, the lack of intermediate conjugate points along the orbits implies the stability of these kinks. However, the stability of these kinks goes beyond linear stability, for these loops are not contractible on the torus and hence these kinks are globally stable.

\section{Conclusions and further comments}

\label{sec:4}

In this paper a rich family of potentials with different number of vacua distributed on the torus has been presented. These models exhibit on one hand singular kinks traveling between vacua and on the other hand families of kinks replicated in the regions delimited by precisely these singular kinks. While the explicit expressions for the singular kinks are available, those of the families are not. However, the orbit equation for these families can be integrated as it has been shown in the cases described above.

Three cases have been chosen for different values of $n_1$ and $n_2$, which in turn determine the number of vacua in the model as well as their position on the torus. Models with eight, four and two vacua have been considered. Notice that there is not a one-to-one correspondence between potentials and number of vacua. Indeed, different combinations of $n_1$ and $n_2$ will correspond to the same number of vacua $|4 n_1 n_2|$. In light of this multiplicity of potentials, only the three mentioned representative cases have been presented.

The spectral analysis of the second order small kink fluctuation operator for every singular kink in our cases reveals their linear stability. On the other hand, the absence of points along the orbits of families of kinks where the flow is undefined makes these families also linearly stable. All the kinks found in the first two cases, singular kinks and members of the families, are topological. However, in Case $3$ non-topological kinks are also found. Indeed, half of the singular kinks and all the families of kinks are non-topological in this last case. As mentioned before, this represents an example where the topology of the target space prevents non-topological kinks from decaying to the vacuum due to the non-contractibility of these loops. Therefore, the global stability of these kinks is guaranteed by topological arguments and they are not only linearly stable but also globally stable.

Finally, as mentioned before the minimum number of vacua that our family of potentials admits is two given the periodicity condition on the potential. Unfortunately, this excludes models with only one vacuum point. The existence of these on the torus would imply that every kink is non-topological, which leaves a gap in this line of work that might be filled in the future.

\section*{Acknowledgements}

The authors thank the Spanish Ministerio de Econom\'{\i}a y Competitividad (MINECO) and the Junta de Castilla y Le\'on for partial financial support under grants PID2020-113406GB-I00 MTM and SA067G19, respectively.

\appendix

\section{Appendix. Differential geometry on the Torus}

Several well-known results about Riemannian geometry of the torus that have been needed along the paper are collected in this Appendix. The notation is adapted to the chosen parametrization (\ref{tor}) to embed the torus in the Euclidean space ${\mathbb R}^3$, i.e. equations (\ref{tor}):
\begin{eqnarray*}
\phi^1(\theta,\varphi) &=& (R+r\sin \theta)\cos \varphi \, ,  \\
\phi^2(\theta,\varphi) &=& r \cos \theta \, ,\\
\phi^3(\theta,\varphi) &=& (R+r\sin \theta)\sin \varphi
\end{eqnarray*}
with: $\theta\in [0,2\pi)$ and $\varphi\in[0,2\pi)$, that convert the constraint (\ref{constraint}):
\[
(R-\sqrt{\phi^1 \phi^1 + \phi^3\phi^3})^2 + \phi^2\phi^2 -r^2=0
\]
to an identity. The Euclidean metric in ${\mathbb R}^3$ is thus reduced to the torus in the form:
\[
ds^2=\left. \left( d\phi^1 d\phi^1 +d\phi^2 d\phi^2 +d\phi^3 d\phi^3\right)\right|_{{\mathbb S}^1\times{\mathbb S}^1} = r^2 d\theta^2+ (R+\sin \theta)^2 d\varphi^2 = g_{11} d\theta^2+ g_{22} d\varphi^2
\]
and this metric tensor gives rise to the following Christoffel symbols:
\begin{eqnarray*}
&&  \Gamma^k_{ij}=\frac{1}{2} g^{kl}\left(\frac{\partial g_{il}}{\partial u^j}+\frac{\partial g_{l j}}{\partial u^i}-\frac{g_{ij}}{\partial u^l}\right) \quad , \quad i,j,k, \cdots=1,2\quad , \quad u^1=\theta \, \, , \, \, u^2=\varphi\\
&&\Gamma^1_{11}=\Gamma^1_{12}=\Gamma^1_{21} = \Gamma^2_{11}= \Gamma^2_{22}=0, \hspace{0.5cm} \Gamma^1_{22}=-\frac{1}{r} \cos\theta (R+r\sin \theta),\hspace{0.5cm} \Gamma^2_{12} =\Gamma^2_{21}= \frac{r\cos \theta}{R+r\sin \theta}
\end{eqnarray*}
where the Einstein convention over repeated indices have been used, and $g^{ij}$ denotes, as usual, the components of the inverse metric tensor. The Riemann tensor, in the case of two-dimensional surfaces, reads:
\begin{eqnarray*}
&& R_{ijkl}=K\left(g_{ik}g_{jl}-g_{il}g_{jk}\right)\\
&& R_{1111}=R_{1112}=R_{1121}=R_{1122}=R_{1211}=R_{1222}=0\\ && R_{2111}=R_{2122}=R_{2211}=R_{2212}= R_{2221}=R_{2222}=0, \\ && R_{1212}=-R_{1221}=\frac{1}{r} \sin \theta (R+r \sin \theta) \hspace{0.5cm},\hspace{0.5cm} -R_{2112}=R_{2121} = \frac{r \sin \theta}{R+r \sin \theta} \, \, .
\end{eqnarray*}
where $K$ stands for the Gaussian curvature. Finally the Ricci tensor components and the scalar curvature are in turn:
\begin{eqnarray*}
&& R_{ij}=R^l_{ilj}=Kg_{ij} \quad , \qquad R=g^{ij}R_{ij}=g^{11}R_{11}+g^{22}R_{22} \\&&
R_{12}=R_{21}=0 \quad , \quad R_{11}=\sin\theta(\frac{R}{r}+\sin\theta) \quad , \quad R_{22}=\frac{\sin\theta}{\frac{R}{r}+\sin\theta} \\ && R=2K= 2 \frac{\sin\theta}{r(R+r\sin\theta)}\quad .
\end{eqnarray*}

\end{document}